\def\kms{km s$^{-1}$\/}
\documentclass{PoS}

\title{Fe II velocity shifts in optical spectra of type 1 AGN}

\ShortTitle{Fe II velocity shifts in optical spectra of type 1 AGN}

\author{\speaker{E.Bon$^{1}$},P. Marziani$^{2}$, M. Berton$^{3,4}$,  N. Bon$^{1}$, R. Antonucci$^{5}$, M. Gaskell$^{6}$,  G. Ferland$^{7}$\\
	$^{1}$Belgrade Observatory, Belgrade, 11060, Serbia; \\
	$^{2}$INAF, Osservatorio Astronomico di Padova, Padova, 35122, Italy; \\
	$^{3}$Finnish Centre for Astronomy with ESO (FINCA), University of Turku, 20014, Finland;\\
	$^{4}$Aalto University Mets\"ahovi Radio Observatory, Kylm\"al\"a, 02540, Finland;\\
	$^{5}$Department of Physics, University of California at Santa Barbara, CA 93106-9530, USA;\\
	$^{6}$Department of Astronomy \& Astrophysics, University of California at Santa Cruz, CA 95064, USA;  \\
	$^{7}$Department of Physics and Astronomy, University of Kentucky, Lexington, KY 40506, USA.

 E-mail: \email{ebon@aob.rs}}

\abstract{Here we present a critical view of practical problems in the analysis of optical Fe II emission lines of Type 1 AGN spectra. Besides the very complex and unclear physical interpretation of the Fe II template shape, there are other issues that might affect the results of Fe II contribution as well (like the S/N, the AGN continuum component modelling, the complex shapes of other broad and narrow emission lines that are blending in the same part of the spectrum, the galactic host stellar component that could lead to mimicking of the Fe II template shape on some parts of spectrum, etc.). In this paper we concentrate particularly on the claims of possibility that in some objects Fe II could be strongly shifted to the red. We examine the effects in the fitting procedure that could artificially lead to such results.
	}

\FullConference{Revisiting narrow-line Seyfert 1 galaxies and their place in the Universe - NLS1 Padova\\
		9-13 April 2018 \\
		Padova Botanical Garden, Italy}

\begin{document}

\section{Introduction}

We examine the question whether the substantial redshifts  of the optical Fe II lines (often > 2000 \kms) that are reported in \cite{huetal08a} are real. 
Sulentic et al. \cite{sulenticetal12} claimed that large Fe II shifts were not real and that Fe II emission could not be well characterized in case Fe II is weak and relatively broad. Subsequent work shall left the issue open. 
Conversely, Ferland et al. (2009) \cite{ferlandetal09} proposed that these shifts could be due to preferentially an
inflowing motion of gas.
But the claimed red shifts are huge (> 2000 km/s). Could they be real?
Could it be that the fitting procedures is artificially causing the apparent shifts?

In order to ascertain the reality of the shifts we performed a dedicated analysis of the spectra indicated by Hu et al. as the best examples of high amplitude redshifts. In the following we describe our selection criteria, our method of analysis, our main results and we discuss some interpretations also outlining the main difficulties in the optical Fe II measurements.

There are several serious problems in the analysis of optical Fe II emission lines of type-1 AGN spectra. The most widely employed technique involves an empirical template based on the Fe II spectrum of the NLSy1 galaxy IZw1\cite{borosongreen92} or on theoretical model template from photo-ionization computations \cite{bruhweilerverner08} .

In this short contribution we will address mainly technical issues of the fitting procedure, leaving to a possible following paper more complete analysis of the interpretation of the Fe II emission (Gaskell et al. in preparation).

{This problem is particularly important for our comprehension of the narrow-line Seyfert 1 (NLS1) phenomenon (Osterbrock \& Pogge 1985)\cite{osterbrockpogge85}. Often considered young AGN (Mathur 2000)\cite{mathur00}, some authors proposed that they are instead sources with a flattened broad-line region observed at very low inclination (Decarli et al. 2008)\cite{Decarli08}. The presence of Fe II red shifts found by Hu et al. (2008), along with the high incidence of blue shifted [O III] in NLS1s (Komossa et al. 2008)\cite{komossaetal08}, produced another hypothesis. Boroson (2011)\cite{boroson11} suggested that NLS1s are low inclination objects with strong outflows, which appear as [O III] blueshift, while the Fe II redshifts are produced by inflowing material moving on the BLR disk. Determining whether or not the Fe II shifts are real is, therefore, of paramount importance to confirm (or disproof) this scenario.}

 \section{Sample selection and Data analysis}
 \label{sec:data}
 
 In order to examine the highest Fe II redshifts,  and avoid unwanted effects of very weak Fe II emission that in lower S/N spectra we 
 selected objects 
 according the results there obtained in \cite{Hu08a}, selecting a subsample of {Fe} {II}   shifted candidates. In order to have significant Fe II emission, we used parameter R$_{\rm Fe}$ that represents the ratio of equivalent widths of Fe II and H$\beta$ broad line (we note that the high R$_{\rm Fe}$ does not necessarily imply strong Fe II emission, since it might result from a week H$\beta$ broad emission line, but with visual inspection of candidates such week H$\beta$ could be  identified and removed from a sample).  
 We therefore selected a subsample with:
 \begin{enumerate}
 \item  Fe II   shift is larger then 800 \kms, and R$_{\rm Fe}$ > 0.8 (in order to select only those objects which have a significant {Fe} {II}  , selected with the aim to avoid spectra with low contribution of {Fe} {II}  , but to have a significant shift;
 \item {Fe}{II}   Shift of {Fe} {II}   > 600 \kms, R$_{\rm Fe}$ > 1 and whose width of {Fe}{II} lines FWHM < 2200 \kms\ that was selected in order to avoid the effects of broadening on the shifts.
 \end{enumerate}

 We modeled the spectra including all components expected to be present, fitted simultaneously. 
 Particular importance has the 
 Fe II template which was assumed to be as provided by Marziani et al (2009) \cite{marzianietal09}. The host galaxy spectrum was computed with a spectral population synthesis program and was included in the simultaneous fit of all components. The broad H$\beta$ line was modeled with two components to be consistent with the procedure obtained in Hu et al. (2008) \cite{huetal08a}. 
 
  We used ULYSS code Koleva et al (2009) \cite{Kol09},\footnote{{The ULySS full spectrum fitting package is available at: http://ulyss.univ-lyon1.fr/}} 
  adopted to fit Sy1 spectra as presented in \cite{bonetal16}. Stellar populations synthesis was generated using Miles spectral library \cite{Miles2006} in order to fit the wavelength range that includes the [O II]$\lambda$ 3727 \AA\ line.The {Fe}{II}   template adopted from \cite{marzianietal09}. Spectra were fitted in the wavelength domain 3700-5700 \AA.

\begin{figure}[ht!]
\begin{center}
	\includegraphics[width=7cm]{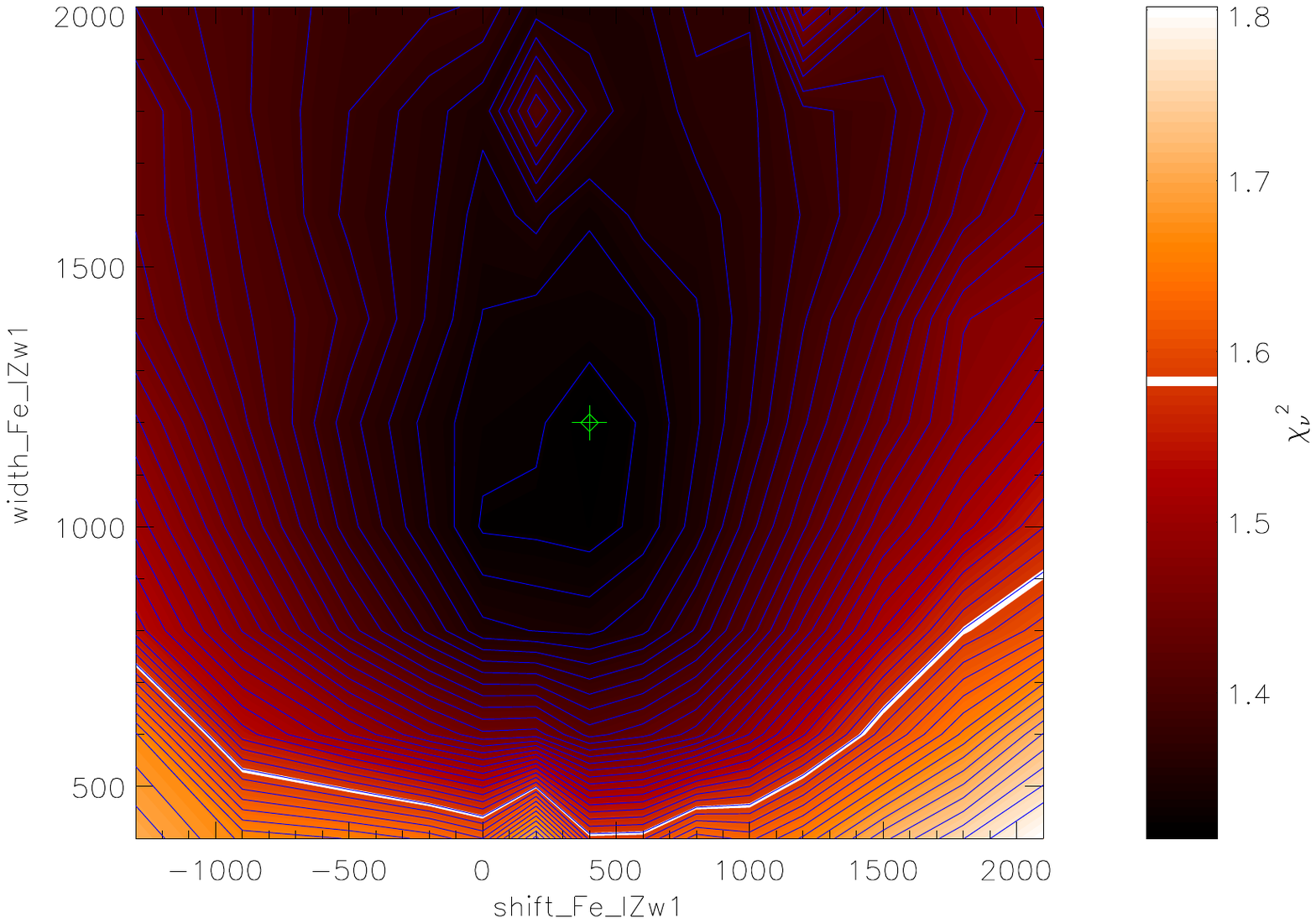}
	\includegraphics[width=7cm]{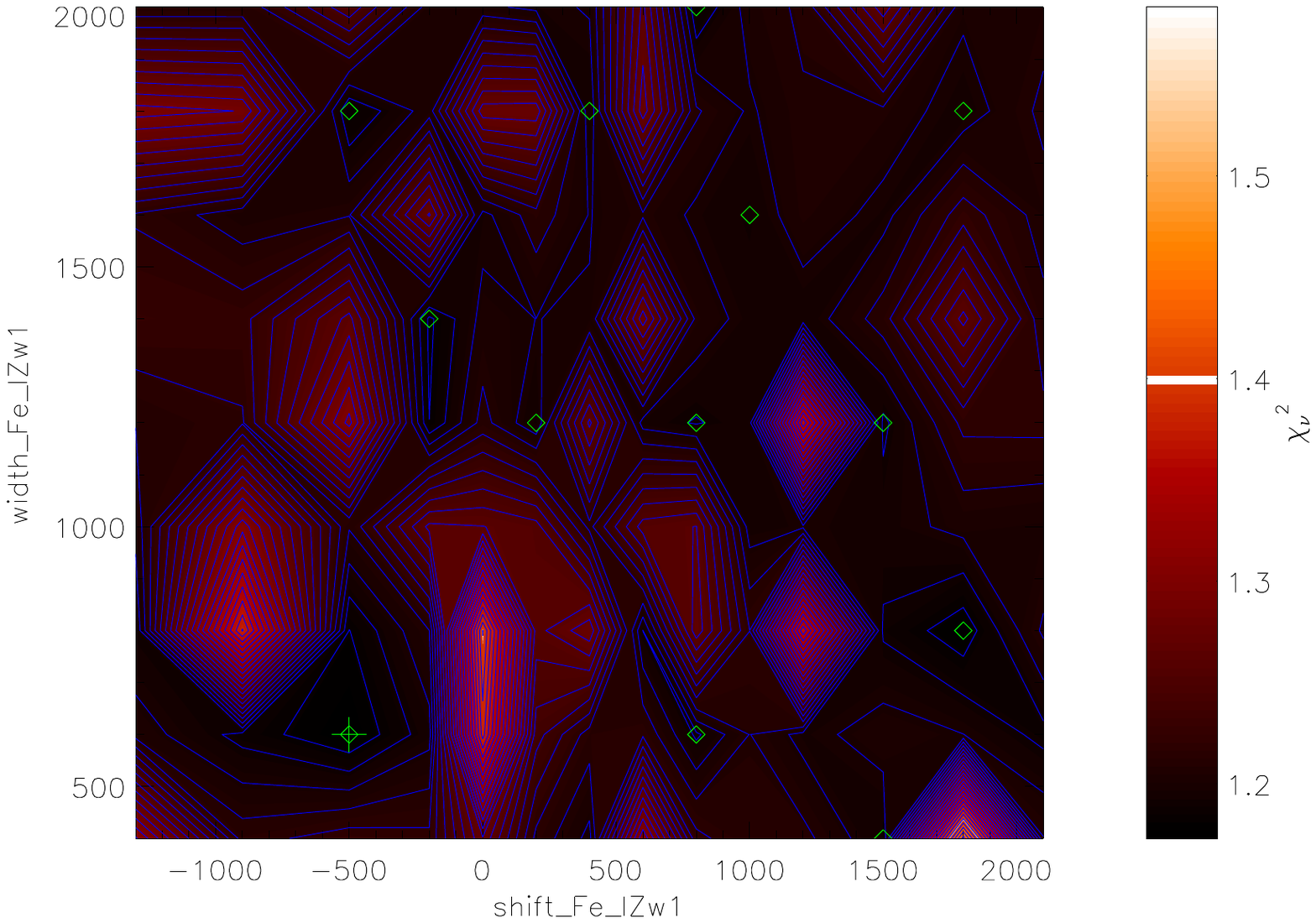}
		\includegraphics[width=8cm]{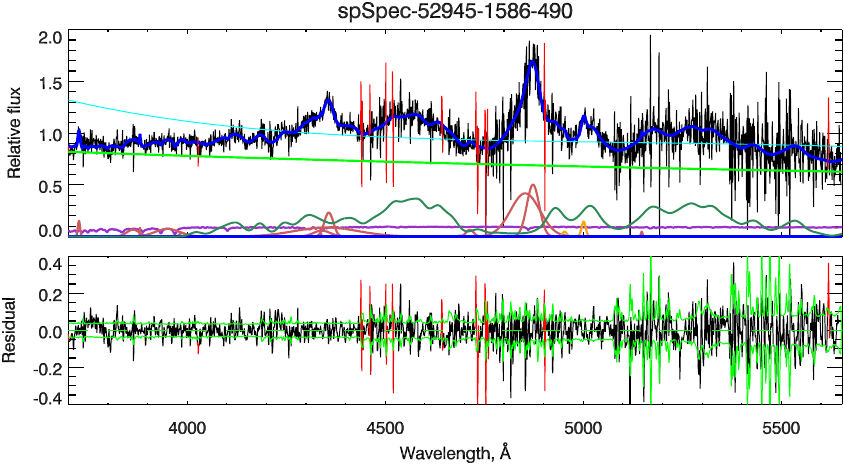}
	\includegraphics[width=8cm]{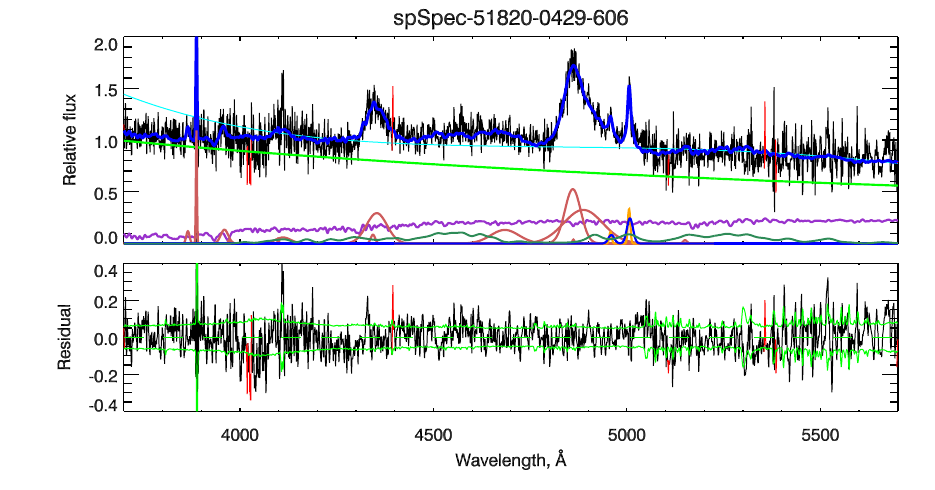}
	
\end{center}
	\caption{Examples of "good" global minimum (SDSSJ162901.43+304401.3, spSpec-52945-1586-490, top left) and "bad" one  (SDSSJ014655.78+135915.4, spSpec-51820-0429-606 top right) on chi square maps of Fe II shift and width parameters (top panel) with corresponding plots of the fitted model for each chi square map (mid and bottom panel). Darker colors correspond to lower value of chi square, with gradient marked with blue iso lines.}
\label{fig:goodbad} 
\end{figure}

We  noticed that the results of fitting in many cases were gathering close to the initial values of the {Fe}{II} shift, indicating degeneracy of some parameters and possible ending of fitting in local minima. Therefore we made low resolution chi square maps using a grid of {Fe}{II}   shift and the {Fe}{II}   widths, since we noticed that the shifts of Fe II were influenced by the lines widths in the fitting procedure, as described in Bon et al. (2016) \cite{bonetal16}. We examined these maps and eliminated the objects showing square maps with many local minima that did not allow clear identification of global minimum (we call it here "bad" maps). In the case of the bad chi square maps there is no defined unique solution for the Fe II template shift and width. An example of "good" and "bad" chi square map with the corresponding fit of each spectrum obtained starting from values of Fe II shift and width from the chi square map, is presented in Fig. \ref{fig:goodbad}. From total of 260 spectra, there were 69 that we selected as "good".
 
Besides the complex and unclear physical interpretation of the Fe II template shape, other issues that might affect estimates of the Fe II contribution are:

	1) S/N ratio,
	 
	2) AGN continuum shape, 
	
	3) complex shapes of other blended broad and narrow line,
	
	4) shape of stellar continuum which could be mistaken for the Fe II template shape in some regions.

Low signal to noise ratio matters because it could lead to artificial broadening  of the Fe II emission lines (we assumed that the broadening of Fe II should be similar to, or smaller value than of broadening of H$\beta$ broad emission line, see for example \cite{Cracco16}). AGN continuum is often uncertain due to contamination by strong broad and blended emission lines. For example H$\beta$ and [O III]4959,5007\AA\ form a blend that is difficult to deconvolve because of the complexity of the [O III] line profile, which is rarely reproducible with a single Gaussian component. The stellar continuum of the underlying host galaxy deserves a special attention. It can mimic the Fe II emission in the spectral range short wards of the  H$\beta$  line (Bon et al. in preparation).

\section{Results}
\label{sec:results}

\begin{figure}[ht!]
	\begin{center}
		
		\includegraphics[width=8cm]{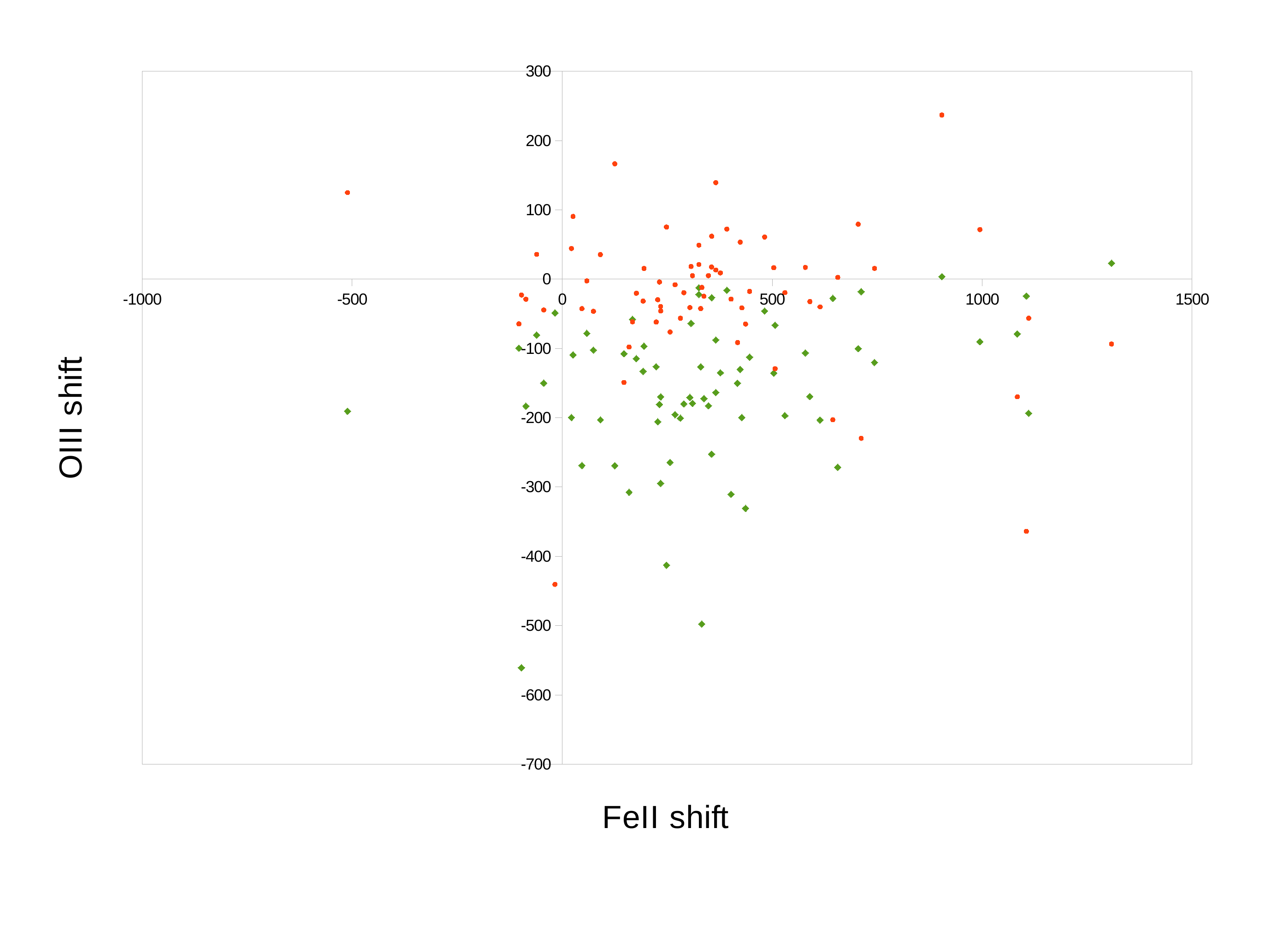}

	\end{center}
	\caption{The shift of Fe II template against the [O III] line component shift of narrow (orange) and semi broad Gaussian component (green).}
	\label{fig:gamma} 
\end{figure}

Using the stellar galactic host component as a kinematic reference frame, we calculated shifts of emission lines.

We find that in general Fe II shifts are correlated with H$\beta$ broad line shifts. The result of correlation coefficient between these two shifts is 0.55 with P=3e-5.
The paired differences of Fe II and   H$\beta$ broad line shifts are modest with the average shift of about 200 \kms.
We were unable the detect shifts as large as those found in Hu et al. (2008)\cite{huetal08a}. We found just few cases above 800 \kms, with the upper limit at 1400 \kms. Objects with Fe II shifts larger than 800 \kms are presented in Table 1.

We found that the analysis of Hu  overestimated the parameter R$_{\rm Fe}$ as well as the shift of the Fe II. We selected sources with criteria on R$_{\rm Fe}$ excluding sources with low Fe II emission. However a posteriori analysis show that the average R$_{\rm Fe}$ is smaller than the lower limit that we used in the selection criteria. This could be the effect of neglecting the host stellar contribution in the analysis of Hu et al. \cite{huetal08a}. A similar effect is noticed by Sniegowska et al. \cite{sniegowskaetal17} in the case of measurements obtained by Shen et al. \cite{shenetal11}.
As for Fe II shifts connection with the orientation hypothesis, suggested in Boroson (2011) \cite{boroson11}, our results do not show that the distribution of Fe II against the [O III] line shifts in reference to the stellar host galactic shift (see, Fig. \ref{fig:gamma}) is similar to the one presented in Fig. 1 of Boroson's manuscript \cite{boroson11}, neither for the narrow or semi broad [O III] component.

\begin{table}[h!]
	\begin{center}
		\caption{Table of objects with Fe II shifts larger then 800 \kms in the kinetic reference frame of the stellar host component}
		\label{tab:table1}
		\begin{tabular}{|l|c|c|c|r|} 
			\hline
			Name & R$_{\rm Fe}$ & H$\beta$ shift & H$\beta$ FWHM & Fe II shift \\
						& & (\kms) & (\kms) & (\kms)\\
						
			\hline
SDSSJ013419.11-084714.5	&	1.48	&	999	&	2772	&	904	\\
SDSSJ141318.96+543202.4	&	0.74	&	643	&	4223	&	995	\\
SDSSJ133450.43+010219.0	&	0.82	&	277	&	4771	&	1084	\\
SDSSJ150226.55+544633.5	&	0.37	&	328	&	4011	&	1105	\\
SDSSJ134503.10+381504.3	&	0.34	&	1092	&	3050	&	1111	\\
SDSSJ080320.30+294548.8	&	0.55	&	562	&	4844	&	1308	\\
		\hline
		\end{tabular}
	\end{center}
\end{table}

\section{Discussion and Conclusions}
\label{sec:discussion}

 Our results confirm the presence of modest shifts among the sources that were identified by Hu et al. (2008) as very shifted sources. The selected subsample with shifts larger than 600 \kms that we analyzed here shows that the shifts are not as high as claimed in Hu et al. (2008). 
 The average shift value is  in agreement with the one computed on the  complete sample of Hu et al (2008). We find that Fe II shifts are  correlated with shifts of H$\beta$ broad component. The value of the difference between shifts of Fe II and H$\beta$ shifts are also in agreement with the average value  obtained for the complete sample in  Hu et al. (2008), indicating a small offset of the Fe II shift of about a 200 \kms. Such small value of redshift difference has multiple interpretations and the issue will be discussed in a future paper. Similar shifts of  Fe II and  H$\beta$ indicate that both emitting  regions are connected physically. The small shift amplitude may make it easier to explain the redshift by a modest radial drift in the outer part of the BLR, as originally suggested by  Ferland et al. (2009) \cite{Ferland09}.



\



\section*{Acknowledgements}

This research is part of the projects 176003 ''Gravitation
and the large scale structure of the Universe'' and 176001 ''Astrophysical
spectroscopy of extragalactic objects'' supported by the Ministry of Education
and Science of the Republic of Serbia.
This research has made use of the NASA/IPAC Extragalactic Database (NED), which is operated by the Jet Propulsion Laboratory, California Institute of Technology, under contract with the National Aeronautics and  Space Administration. Funding for the Sloan Digital Sky Survey has been provided by the Alfred P. Sloan Foundation and the U.S. Department of Energy Office of Science. The SDSS web site is \texttt{http://www.sdss.org}. SDSS-III is managed by the Astrophysical Research Consortium for the Participating Institutions of the SDSS-III Collaboration including the University of Arizona, the Brazilian Participation Group, Brookhaven National Laboratory, Carnegie Mellon University, University of Florida, the French Participation Group, the German Participation Group, Harvard University, the Instituto de Astrofisica de Canarias, the Michigan State/Notre Dame/JINA Participation Group, Johns Hopkins University, Lawrence Berkeley National Laboratory, Max Planck Institute for Astrophysics, Max Planck Institute for Extraterrestrial Physics, New Mexico State University, University of Portsmouth, Princeton University, the Spanish Participation Group, University of Tokyo, University of Utah, Vanderbilt University, University of Virginia, University of Washington, and Yale University.

This conference has been organized with the support of the
Department of Physics and Astronomy ``Galileo Galilei'', the 
University of Padova, the National Institute of Astrophysics 
INAF, the Padova Planetarium, and the RadioNet consortium. 
RadioNet has received funding from the European Union's
Horizon 2020 research and innovation programme under 
grant agreement No~730562. 

\bibliographystyle{JHEP}
\bibliography{biblioletter4}


\end{document}